\documentclass{article}
\usepackage{natbib}
\usepackage{psfig,graphicx,epsfig}

\def\gsim{\;\lower4pt\hbox{${\buildrel\displaystyle >\over\sim}$}\,}
\def\lsim{\;\lower4pt\hbox{${\buildrel\displaystyle <\over\sim}$}\,}

\def \xmm {{\em XMM-Newton}}

\def \hcm {\hbox {\ifmmode $ atom cm$^{-2}\else atom cm$^{-2}$\fi}}

\def\approxgt{\mathrel{\hbox{\rlap{\lower.55ex \hbox {$\sim$}}
        \kern-.3em \raise.4ex \hbox{$>$}}}}
\def\approxlt{\mathrel{\hbox{\rlap{\lower.55ex \hbox {$\sim$}}
        \kern-.3em \raise.4ex \hbox{$<$}}}}

\def\lsim{\;\raise0.3ex\hbox{$<$\kern-0.75em\raise-1.1ex\hbox{$\sim$}}\;}
\def\gsim{\;\raise0.3ex\hbox{$>$\kern-0.75em\raise-1.1ex\hbox{$\sim$}}\;}

\def\beq{\begin{equation}}
\def\enq{\end{equation}}
\def\begar{\begin{eqnarray}}
\def\endar{\end{eqnarray}}
\def\mathnew{\mathsurround=0pt}
\def\simov#1#2{\lower .5pt\vbox{\baselineskip0pt \lineskip-.5pt
        \ialign{$\mathnew#1\hfil##\hfil$\crcr#2\crcr\sim\crcr}}}

\def \xmm {{\it XMM-Newton}}
\def \srca {IC443}
\def \srcb {G166.0+4.3}

\begin{document}


\LARGE
\begin{center}

{\bf On the metal abundances inside mixed-morphology supernova remnants: the case of \srca\  and \srcb}

\end{center}

\vspace{0.2cm}


\Large
{F. Bocchino(1)
M. Miceli(2)
E. Troja(3,4)
}

\vspace{0.5cm}

\large
1 - INAF-Osservatorio Astronomico di Palermo, Piazza del Parlamento 1,
       90134 Palermo, Italy \\

2 - Consorzio COMETA, Via S. Sofia 64, 95123 Catania, Italy \\

3 - INAF - Istituto di Astrofisica Spaziale e Fisica Cosmica, Sezione di Palermo, via Ugo la Malfa 153, 90146 Palermo \\

\normalsize

\abstract
  {Recent developments on the study of mixed morphology supernova remnants
  (MMSNRs) have revealed the presence of metal rich X-ray emitting plasma
  inside a fraction of these remnant, a feature not properly addressed
  by traditional models for these objects.}
  {Radial profiles of thermodynamical and chemical parameters are needed
  for a fruitful comparison of data and model of MMSNRs, but these are
  available only in a few cases.}
  {We analyze XMM-Newton data of two MMSNRs, namely \srca\  and \srcb,
  previously known to have solar metal abundances, and we perform spatially
  resolved spectral analysis of the X-ray emission.}
  {{    We detected enhanced abundances of Ne, Mg and Si in the hard
  X-ray bright peak in the north of \srca, and of S in the outer regions
  of \srcb. The metal abundances are not distributed uniformly in both
  remnants.  The evaporating clouds model and the radiative SNR model
  fail to reproduce consistently all the observational results.}}
  {We suggest that further deep X-ray observations of MMSNRs may reveal
  more metal rich objects. More detailed models which include ISM-ejecta
  mixing are needed to explain the nature of this growing subclass of
  MMSNRs. }



\section{Introduction}

Mixed-morphology supernova remnants (MMSNRs) have been traditionally
defined as remnant having a shell morphology in radio and a centrally
peaked morphology in the X-ray band characterized by a thermal spectrum 
(\citealt{rp98}). The origin of
this atypical morphology is controversial. Traditional models 
mostly relies on the effects of thermal conduction (e.g. \citealt{csm99},
\citealt{scm99}, \citealt{wl91}), but there are also models which invoke
projection effects (\citealt{pet01}). \citet{tbh06} tried to
refine previous analytical models by using a set of hydrodynamical
simulations aimed at reproducing the morphology observed in MMSNRs.

In parallel to these theoretical efforts, there is a growing
observational interest around MMSNRs, which has led in a few cases to
useful comparison with models. \citet{ssh02} favor the evaporating clouds
\citet{wl91} model for G290.1-0.8 (rather then the radiative model of
\citealt{scm99}), but problems remain in the comparison of the surface
brightness profiles. \citet{ls06} also find a reasonable agreement
between both models and X-ray observations of CTB1 and HB21, but they
point out that the predicted central density are below the derived
densities of the emiting plasma. \citet{rb02}, instead, argue that the
\citet{wl91} evaporating cloud model is at odds with the presence of a
strong radiative shock front in the MMSNR W28, but also the radiative
model is at odd with the strong thermal variation found in this remnant
(see also \citealt{che99} for a critical approach to the evaporating
cloud model of MMSNRs).

In addition, several authors have pointed out that part of the
X-ray emission of MMSNRs may be composed by thermal plasma with high
metal abundances, as those observed in stellar ejecta inside young
historical SNRs (e.g. \citealt{skp04}, \citealt{css08}), {    or in other
Galactic (e.g. Vela SNR, \citealt{mbr08}; Cygnus Loop, \citealt{ktm08};
Puppis A, \citealt{hpf08}) and Magellanic Cloud SNRs (e.g. 0103-72.6,
\citealt{phb03}; N49B, \citealt{phs03}).} This important result was
addressed and summarized by \citet{ls06}, which report a compilation
of 26 MMSNRs, 10 of which show some sign of enhanced metal abundances
in the X-ray spectrum.  It is important to stress at this point that
the presence of enhanced metal abundances inside MMSNRs has not been
addressed in detail yet in the models developed for this sub-class of
remnants. In general, published models with thermal conduction do not take
into account the mixing of ejecta with the circumstellar and interstellar
medium, so the emerging class of metal rich MMSNRs is still not properly
understood. This is also the reason why the comparison between models
and observation has mostly focused only on morphological issues, with
the general comment that the presence of the additional ejecta component
may reconcile the discrepancy between model and observed profile (like
in the case of W44, \citealt{skp04}).

{    The mixed morphology category of SNRs (and especially the emerging subclass
of metal-rich MMSNRs) can be better understood if we have a large
sample of objects with well studied characteristics.  In particular,
the accurate determination of density, temperature and abundances
profile for MMSNRs will allow us to perform an accurate comparison with
both traditional models and a new magnetohydrodynamical (MHD) model of
shocked ISM and ejecta inside MMSNRs, subject
to a forthcoming publication.  

We have undertaken a systematic review of
the MMSNRs listed in \citet{ls06} with the primary goal to measure 
the abundances in the inner region of the remnants and
other X-ray properties.  In this paper, we review the XMM-Newton archive
observations of two bright MMSNRs, \srca\  and \srcb, with the aim of
deriving the distribution of temperature and metal abundances inside
these remnants. Both our targets are mentioned in the list of \citet{ls06}
as MMSNRs with standard metal abundances. However, \citet{tbr06} and
\citet{tbm08} have studied in detail the X-ray emission of \srca, and
they have found enhanced metal abundances in a remarkable ring feature
around the pulsar wind nebula, in the southern part of the remnant, attributed to
emission from shocked ejecta.  
The compilation of MMSNRs with enhanced metal abundances
needs therefore to be updated with new results from deep
X-ray observations of MMSNRs. On the light of these new
observational results, traditional models should be
further verified and revised as necessary.

The paper is organized as follows: in Sect. 2 and 3 we present the
results obtained form XMM-Newton archive observations of \srca\  and
\srcb, respectively. In Sect. 4, we present the comparison of our
results with the standard models for MMSNRs, namely the evaporating
clouds \citet{wl91} model and the radiative SNR model of \citet{csm99}. In
Sect. 5, we present the conclusion of our work.}

\section{\srca}

\begin{figure}
  \centerline{\psfig{file=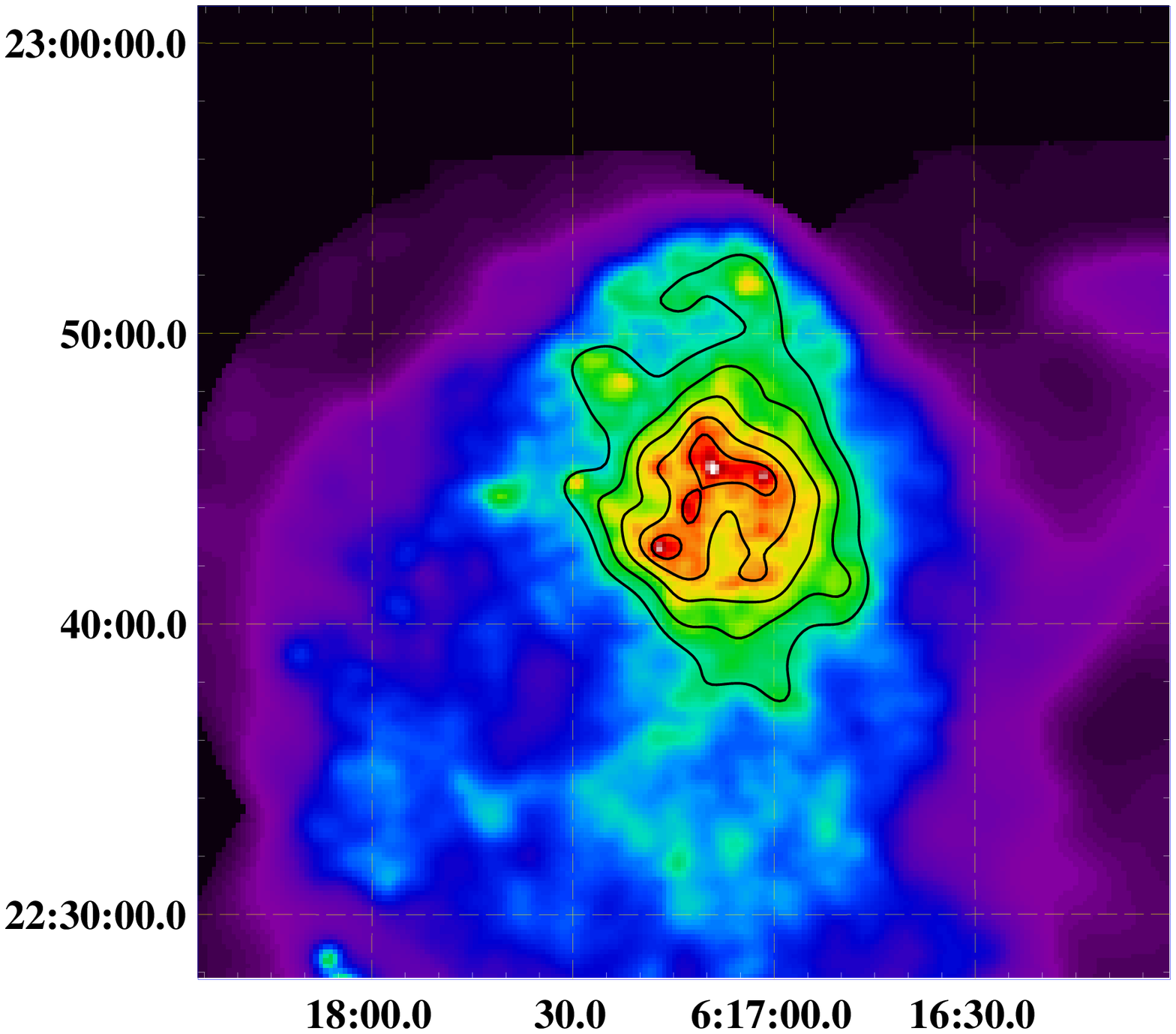,width=8.0cm}}
  \hspace{1.8cm}\psfig{file=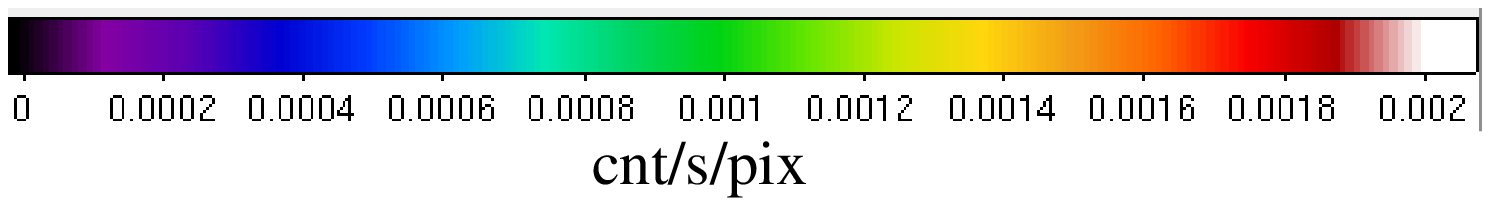,width=6.5cm}
  \centerline{\psfig{file=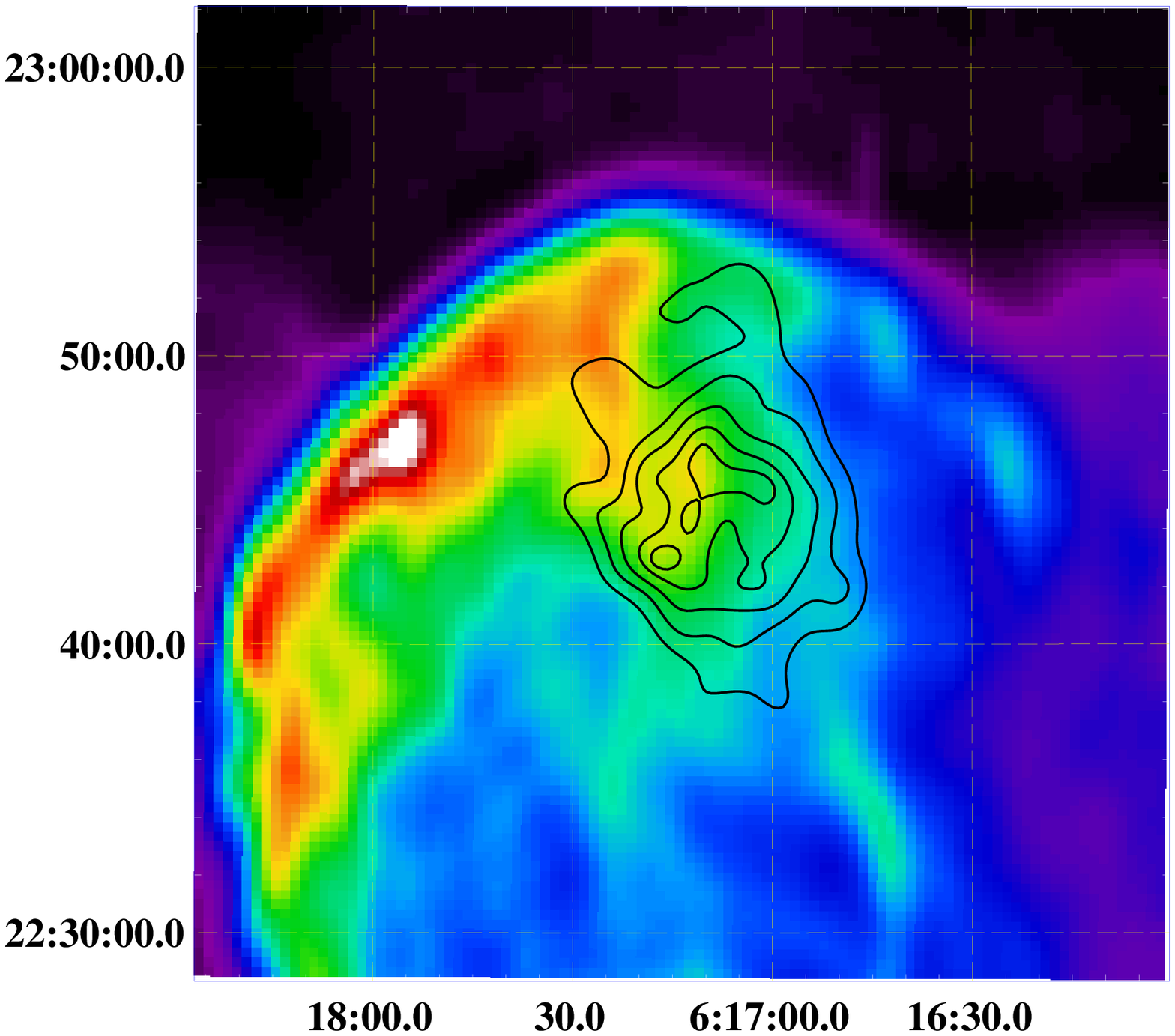,width=8.0cm}}

  \hspace{1.8cm}\psfig{file=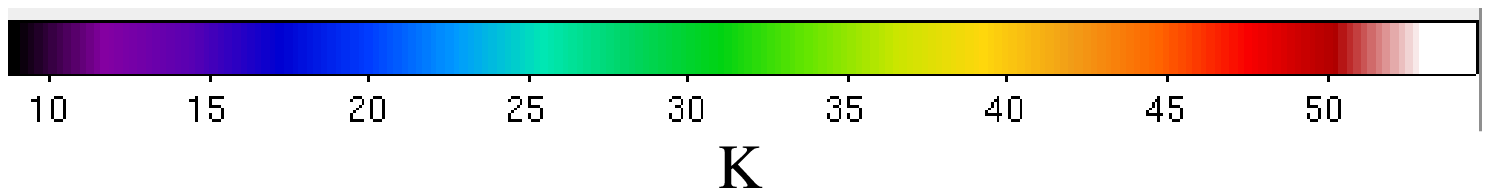,width=6.5cm}

  \caption {{\em Top:} XMM-Newton EPC mosaic of the northern region of
  the IC443 supernova remnant in the 1.4--5.0 keV band. The image is
  background subtracted and vignetting corrected. The pixel size is 10
  arcsec. Five contours levels at 8, 10, 12, 14 and $16\times 10^{-4}$
  cnt s$^{-1}$ pix$^{-1}$ are overlaid. These contours have been used
  as boundaries for the 5 spectral extraction regions discussed in the
  text. {\em Bottom:} same region at 1420 MHz (total intensity map) 
  with X-ray contours overlaid 
  (adapted from \protect\citealt{lea04}).
  }
  \label{ic443}
\end{figure}

\srca\  has been observed in several bands and extensively studied in
the literature and, for this reason, it is a case study for the interaction between SNR
shocks and molecular clouds. Recent X-ray studies includes \citet{tbm08},
\citet{bku08}, \citet{tbr06}, while for infrared and radio studies see
\citet{nhg08}, \citet{bku08}, \citet{lky08}, \citet{raa07}, \citet{lea04}
and references therein. 
The remnant has been classified as mixed-morphology by \citet{rp98},
based on observations with the ROSAT satellite. \citet{tbr06} and
\citet{tbm08} confirm the centrally peaked morphology between 0.5 and 5
keV, with a bright X-ray peak near $06^h17^m09^s$ $+22^\circ 45^m$ in the
1.4--5 keV (Fig. \ref{ic443}).  However, they note that, at very soft energies (0.3--0.5
keV), the remnant shows an incomplete shell extending from N to E,
corresponding to one of the region of interaction with a large cloud.
They have also investigated the spatial distribution of the metals inside
this remnant, using the technique of the equivalent width images. According
to them, the distribution of Si and S is far from being homogeneous. In
particular, a ring of metal rich plasma surrounding the pulsar wind
nebula is evident in the equivalent width images of these two elements
(see their Fig. 3). Moreover, the northern region of the remnant,
coincident with the maximum of the hard X-ray image (Fig. \ref{ic443}),
is also characterized by a large value of the Si and S equivalent width,
suggesting higher metal abundances with respect to surrounding regions. In
the following, we focus on this region.

The data we use in this work are part Calibration and Performance
Verification phase of the \xmm\ satellite (\citealt{jla01}), for a
total of 6 observations of \srca. In addition, we have used the public
archive observation 0301960101. The datasets we have used are the same
datasets used by \citet{tbm08}. We screened both PN and MOS data using
the algorithm suggested by \citet{sk07}.

The results are reported in Fig. \ref{ic443fits}.
In order to study the profile of the thermodynamical and chemical
parameters of this remnant, we have chosen 5 spectral regions defined
in terms of contours of iso-surface brightness, centered in the bright
northern region, as shown in Fig. \ref{ic443}.  For each region, we
have computed a mean distance from the X-ray peak as the average of
the distance of all the pixel in that region. XMM-Newton PN and MOS
spectra have been extracted using the software SAS 7.1 and corrected for
detector non-uniformity using the task {\em evigweight}.  \citet{tbr06}
have shown that the X-ray emission in the \srca\  interior is composed
by two thermal components, the soft one with temperature in the range
0.3--0.7 keV, in non-equilibrium of Ionization (NEI), and the second
one with $kT=1.1-1.8$ keV, in equilibrium of ionization (CIE). The soft
component has been associated with the shocked interstellar material,
while the hot component to the shocked stellar ejecta. {    Therefore,
we have used the same emission models, namely the {\sc mekal} model
(\citealt{mgv85}) and the {\sc VNEI} (\citealt{blr01})
model of XSPEC v11.3.2 (\citealt{arn96}).
We have fixed the chemical abundances of
the soft component to solar values, and left the abundances of the hot
component free to vary. The adopted models represent an approximation of
the true conditions inside this complex remnant. However, this approach
allows us to measure quantitatively the emission measure weighted
temperature and metal abundances for each component in each spectrum.
Moreover, by using the region layout shown in Fig. \ref{ic443}, we may derive
meaningful constrain to any trend of the measured quantities related to
the radial distance of the regions.}

\begin{figure}
  \centerline{\psfig{file=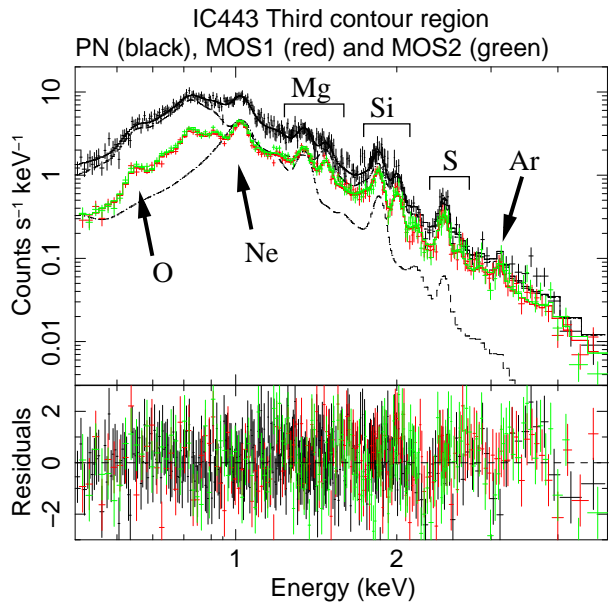,width=8.0cm}}

  \caption {XMM-Newton EPIC spectrum of the third contour region
  in Fig. \protect\ref{ic443} (average distance $2.8^\prime$ in
  Fig. \protect\ref{ic443fits}). The best fit two thermal component model
  is shown as continuous line (with residual in lower panel), while
  individual components are shown in dashed black lines for the PN spectrum 
  only. Emission lines from O
  (0.6--0.8 keV), Ne ($\sim 1$ keV), Mg ($\sim 1.4$ keV), Si (1.8--2 keV),
  S ($\sim 2.5$ keV) and Ar ($\sim 3$ keV) are visible.
  }
  \label{mm3spec}
\end{figure}

In Fig. \ref{mm3spec}, we show an example
of spectrum, while the complete results of the spectral analysis
are reported in Fig. \ref{ic443fits} for the 5 spectral
regions, displayed in order of increasing distance from the putative
center. The temperature does not show large variations, indicating
that thermal conduction must be efficient. The innermost regions are
characterized by a higher metal abundances for Ne, Mg and Si. In these
cases, the profiles show a decreasing trend with increasing distance form
the center. While Ne and Mg retain over-solar of abundances in most
regions, Si is above 1 only in the two innermost regions. On
the other hand, S and Fe seems to have more uniform abundances,
about 1.0 and 0.3 respectively. The thermodynamical parameters of both the
components are in general agreement with the findings of \citet{tbr06}.
{     The abundances patterns are generally consistent with the ones
found by \citet{tbm08} in the south region of the remnant, but smaller
in value if compared to the remarkable ejecta ring surrounding the pulsar
wind bebula found by them. \citet{tbm08} have also compared the Mg/Si,
S/Si and Fe/Si ratios of the ring with the nucleosynthesis model of
core-collapse (\citealt{ww95}) and Type Ia (\citealt{bbb03}) supernovae,
finding more agreement with the former.  We have verified that Mg/Si,
S/Si and Fe/Si ratios found by us in the innermost spectral region in
Fig. \ref{ic443} are in agreement with those of the ring region, thus we
concluded that our results also support a Type II progenitor for IC443.}

\begin{figure}
  \hspace{1.0cm}{\psfig{file=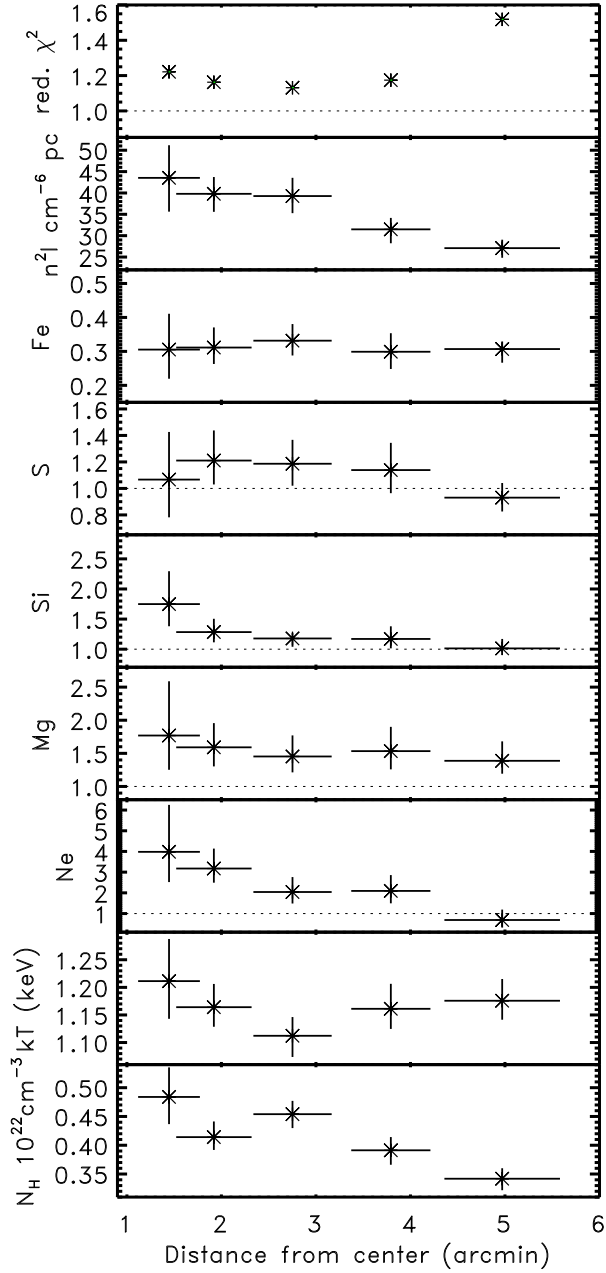,width=8.0cm}}
  \caption {Results of the spectral fittings to the hard thermal
  component obtained in the 5 iso-brightness regions in \srca. See
  Fig. \protect\ref{ic443} for the position of the contours levels
  used for the region definition. Abundances are relative to solar
  values. The abscissa contains the average distance from the center
  (horizontal error bars are standard deviations). Vertical error bars
  are $2\sigma$ uncertainties.  The normalization of the spectrum is
  expressed in terms of $n^2l$, where $n$ is the plasma density and $l$
  is its extension along the line of sight.}

  \label{ic443fits}
\end{figure}

\section{\srcb\ (VRO 42.05.01)}

\srcb\  is a supernova remnant in the anti-center direction whose
X-ray and radio morphology has drawn considerable interest in the
literature. \citet{bg94} and \citet{gb97} have reported on the ROSAT
and ASCA X-ray observations of this remnant, finding a centrally
peaked morphology which seems to be totally enclosed within the radio
emission of this object. The latter, in turn, has a limb brightened
double shell morphology, with a small diameter half-shell at NE (about
30', a.k.a. the shell) and a larger incomplete shell at SW (a.k.a. the
wing, \citealt{ppl85}, \citealt{plr87}, \citealt{lt05} and references therein).

\srcb\  was observed by \xmm\ on 21 February
2003 (ID 0145500101) and on 23 February 2003 (ID
0145500201). The first observation was pointed at the western part of the
remnant (the ``wing''), for a total PN exposure after
the screening procedure of 5.0 ks, while the second one was pointed to the
small incomplete shell at the east. This latter pointing was severely
contaminated by high energy proton flares and the net PN exposure time
is only 1 ks, and we have not used it for spectral analysis.

\begin{figure}
  \centerline{\psfig{file=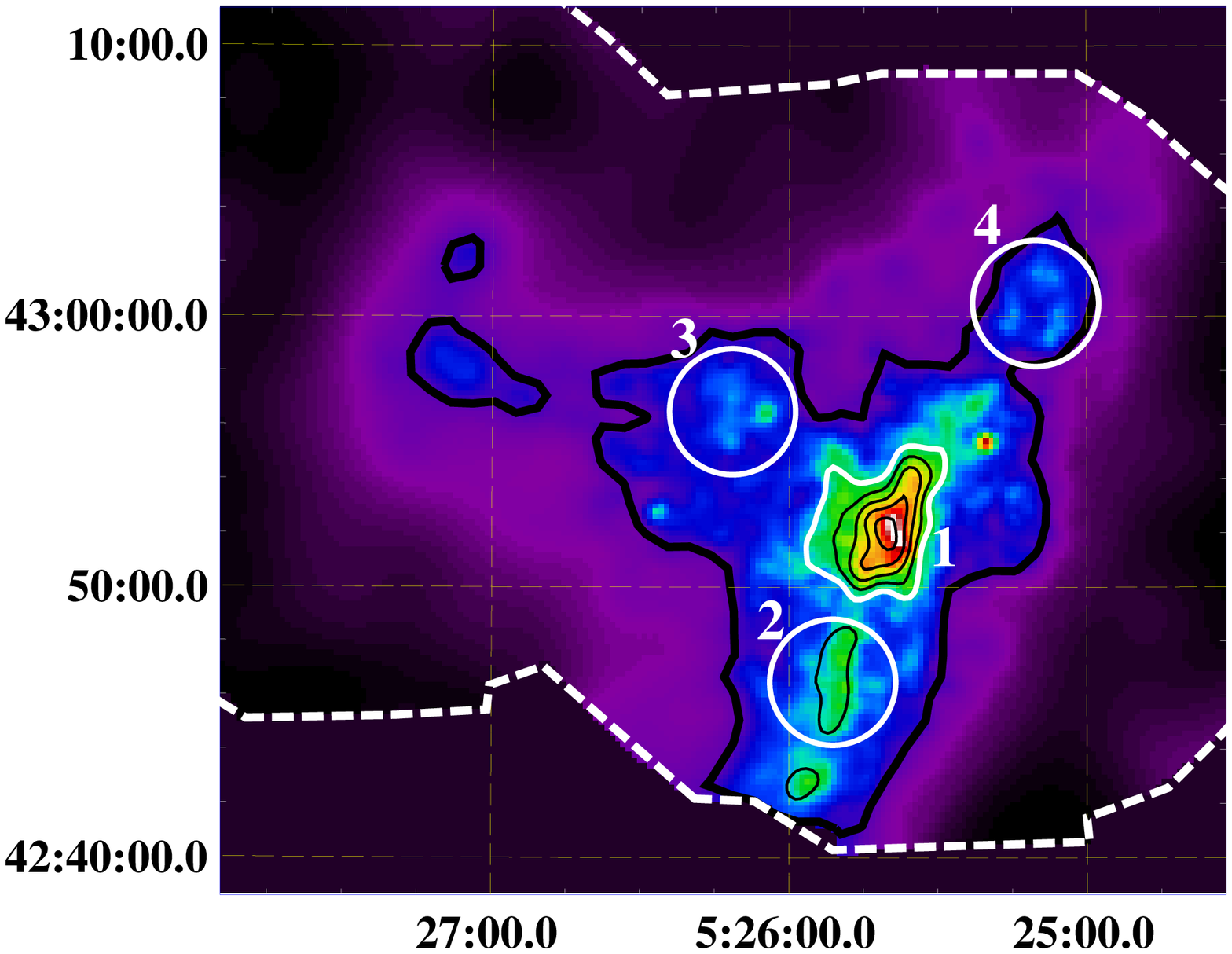,width=8.0cm}}
  \hspace{1.9cm}{\psfig{file=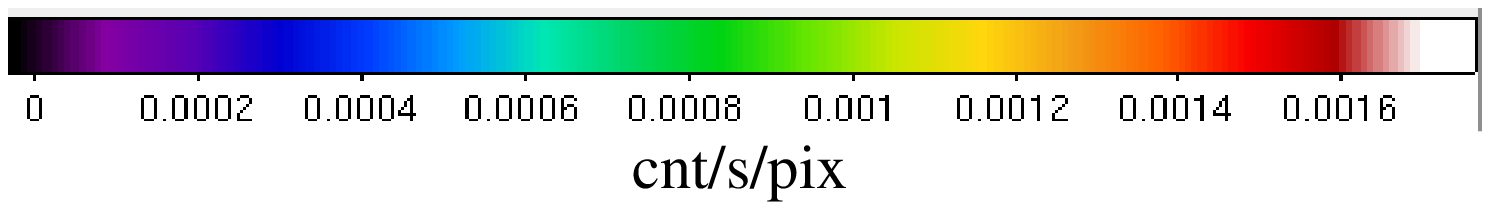,width=6.4cm}}
  \centerline{\psfig{file=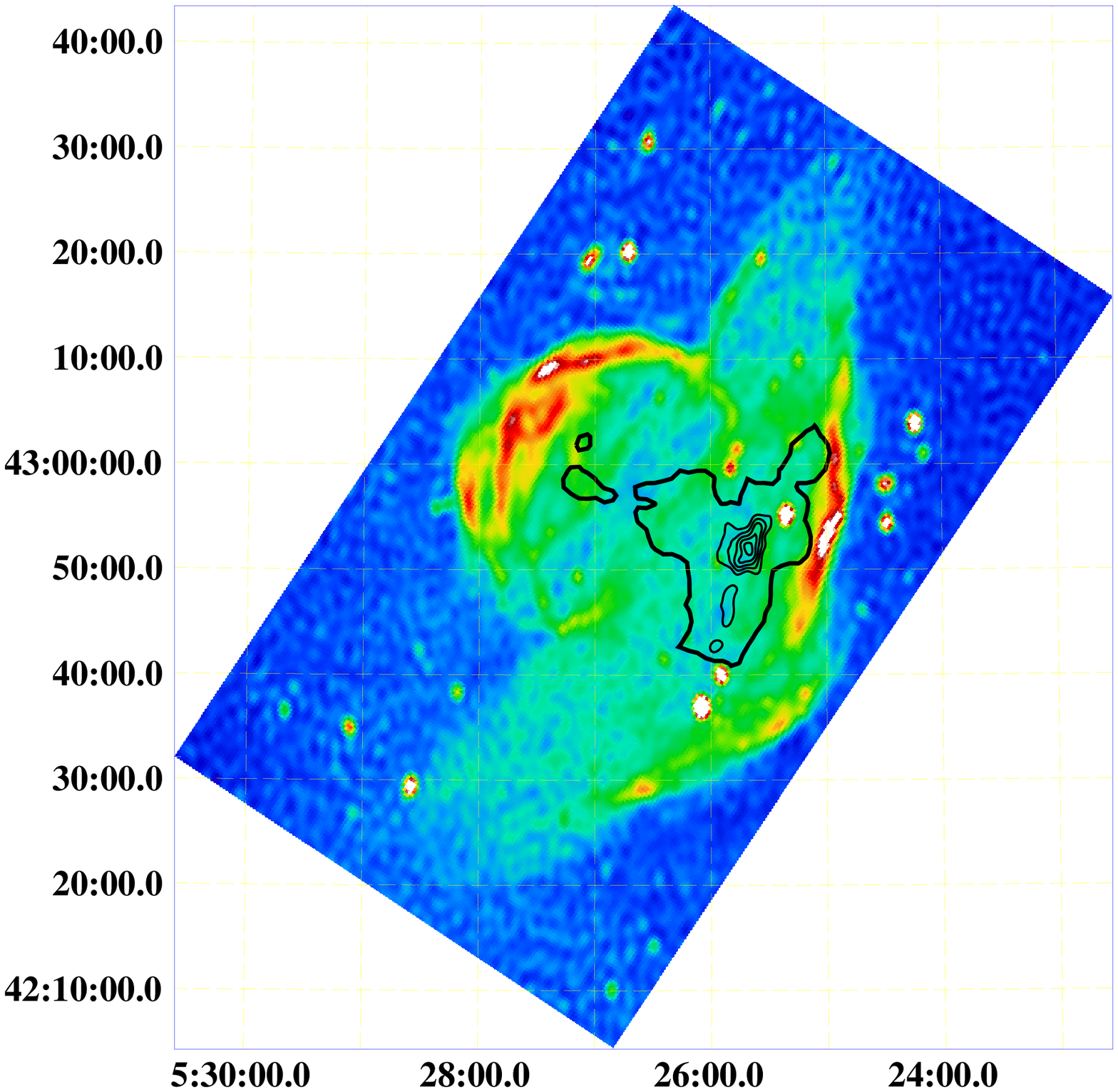,width=8.0cm}}

  \hspace{1.7cm}{\psfig{file=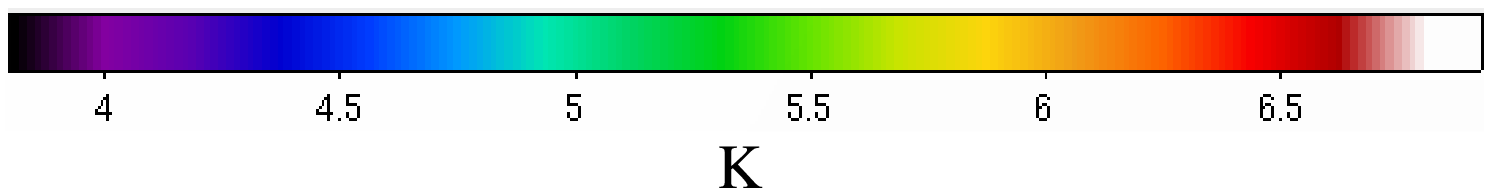,width=6.7cm}}

  \caption {{\em Top:} \xmm\  EPIC images of \srcb\  in the 0.3--5 keV
  background subtracted and vignetting corrected. Black thick contour is
  at $0.2\times 10^{-3}$ cnt s$^{-1}$ pix$^{-1}$ and black thin contours
  correspond to 0.6, 0.8, 1.0, 1.2 and $1.5\times 10^{-3}$ cnt s$^{-1}$
  pix$^{-1}$ (1 pixel = $12^{\prime\prime} \times 12^{\prime\prime}$).
  {    Numbers mark the spectral extraction region used in the text (Region
  1 is defined by the outermost thin surface brightness contour).} The
  \xmm\  EPIC field of view is marked in white (two observations combined).
{\em Bottom} 1420 MHz CGPS DRAO radio image of \srcb\  (adapted from
\protect\citealt{lt05}). X-ray contours of the top panel are shown
in black.
  }

  \label{vro}
\end{figure}

A mosaic map of this remnant in the band 0.3--5.0 keV, obtained using
the two \xmm\  archive observations, is shown in Figure \ref{vro}, which
also include the 1420 MHz radio map of the Canadian Galactic Plane Survey
(\citealt{tgp03}). The X-ray morphology derived by previous observations
is confirmed, and the bright region (named ``western bright knot'' by
\citealt{bg94}) is resolved for the first time, showing an elongation
in the NW-SE direction. On the light of its radio and X-ray morphology,
this remnant may be considered part of the sub-class of mixed-morphology
(MM) SNRs introduced by \citet{rp98}. The adopted explanation of the
peculiar radio morphology dates back to \citet{plr87} and involves the
explosion in a moderately dense medium followed by the breaking out of
the shock in a rarefied hot tunnel, followed in turn by the interaction
with again a denser medium. \citet{gb97} have made the only measurements
of metal abundances in the X-ray spectrum of \srcb, finding a substantial
underabundance of Mg, Si and Fe in the western bright knot and in the
SE and NE of the wing.

\begin{figure}
  \centerline{\psfig{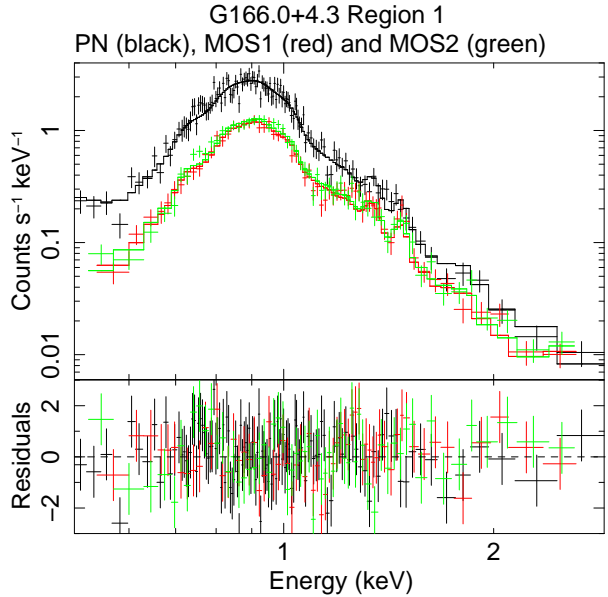}}

  \caption {XMM-Newton EPIC spectrum of the \srcb\  Region 1 in
  Fig. \protect\ref{vro}. The best fit thermal model is shown as continuous
  line (with residual in lower panel).}

  \label{sc15spec}
\end{figure}

\begin{figure}
  \centerline{\psfig{file=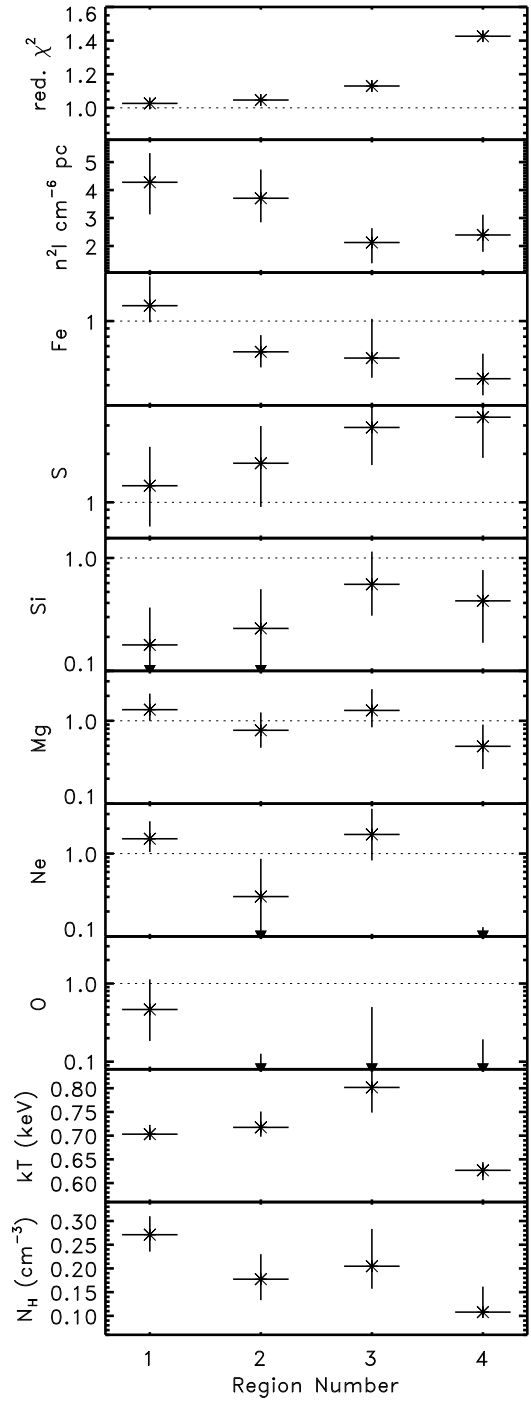,width=7.0cm}}
  \caption {Spectral fitting results of regions 1--4 in \srcb. See
  Fig. \protect\ref{vro} for the position of the regions. Black arrows
  indicates a lower limit of 0.  }

  \label{vrofits}
\end{figure}

In order to study the abundances in the centrally peaked region of
\srcb, we tried to characterize the thermal properties in the
plasma of this remnant by selecting 4 interesting regions for spectral
analysis, which are shown in Fig. \ref{vro}. Regions 1,2 and 3 are
similar to region 1, 2 and 3 of \citet{gb97}, but they are considerably
smaller and pointed to bright features, and therefore they are much
less affected by background (also considering the smaller \xmm\  PSF
compared to the ASCA one). In each region, we fitted the X-ray spectra
with a \citet{mgv85} spectral model modified by interstellar absorption
and letting the abundances of O, Ne, Mg, Si, S and Fe free to vary. In
Fig. \ref{sc15spec}, we show, as an example, the spectrum of region 1. 
The results of the spectral analysis are presented in Fig. \ref{vrofits},
and indicate low variation of temperature (up to 10\%) and large variation
of interstellar absorption (up to 50\%). The variation patterns seems
to be more in agreement with those found by \citet{bg94} than those
of \citet{gb97}. {    The regions seems to be characterized by an
underabundance of O and Si and an overabundance of S, while Ne, Mg and
Fe seems to be around solar values. Our Mg, Si and Fe abundances are in
agreement with \citet{gb97}, who also found a value of $\lsim 1$ for
the abundances of these elements.} A thermal model in Non-Equilibrium of
Ionization do not improve the fits in the regions, giving a ionization
time $\tau \gsim 10^{12}$ s cm$^{-3}$ and abundances very similar to
the one we obtained with the equilibrium model.

We have investigated if there is a radial gradient of metal abundances
variations inside the bright region 1 of the remnant (the ``western
bright knot''), in a manner similar to what we have done in Sect. 2 for
IC443, {    but, given the low counting statistics, the results are
not constraining.}

\section{Discussion}

{    \srca\  and \srcb\  were listed in the \citet{ls06} work as standard
abundances MMSNRs. However, the spatially resolved spectral analysis we
have done on the bright X-ray central regions of IC443 has
pointed out that this object belong to the subclass of metal rich
mixed morphology remnant. In the case of \srcb\  there is no compelling
evidence for enhanced metal abundances apart from sulfur in the outer
regions.  The short exposure time of the XMM-Newton observation of the
latter has prevented us to study possible trends of metal abundances in
the core of this remnant. Our results suggest that it is important to
review the list of MMSNRs of \citet{ls06} to properly address the issue
of the chemical abundances in all the known remnant of this kind. Other
MMSNRs may be erroneously cataloged as standard abundances objects. }

It is not straightforward to understand if the high metal abundances
seen in some of the MMSNRs is a general property, or if represent just
an evolutionary phase of the life of these objects, or it depends on
the initial remnant composition.  From the theoretical point of view,
the evaporating cloud model of \citet{wl91} and the radiative SNR model
\citet{csm99}, traditionally used to explain the centrally peaked thermal
X-ray morphologies, can be of limited help in understanding the subclass
of the metal rich MMSNRs, since they do not consider in detail the
mixing between ejecta and shocked material inside the remnants.  However,
these model have been often used to interpret the X-ray emission of some
MMSNRs (e.g. \citealt{bb03c}, \citealt{ls06}), so it is worth to perform
a detailed comparison with our results to underline their limitations
and to drive the exploration of new ones.

\begin{figure}
  \centerline{\psfig{file=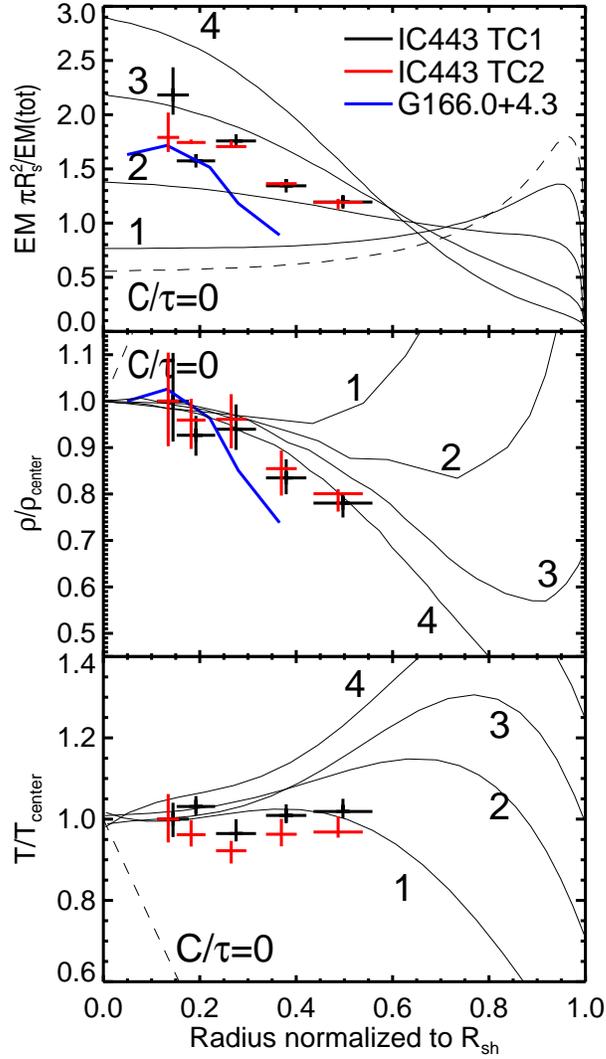,width=8.0cm}}
  \caption{Comparison between the \citet{wl91} evaporating cloud
  model and the observed profiles in \srca\  and \srcb. {\em Top:}
  Normalized emission measure (see text for explanation). {\em Middle:}
  Density normalized to central value. {\em Bottom:} post-shock
  temperature normalized to central value. Black and red crosses are
  IC443 low and high temperature component data with uncertainties,
  respectively; blue curves are \srcb\ data. Black curves are
  \citet{wl91} profiles for models with $C/\tau=0$ (dashed), 1, 2, 3
  and 4. The distance from the center is normalized to the shell radius
  (assumed $10^\prime$ for \srca\  and $\sim 6.5^\prime$ for \srcb,
  see text for details).}

  \label{cfrwl91}
\end{figure}

We first consider the evaporating cloud model of \citet{wl91}, in which
the X-ray central enhancement is due to the material ablated by shock
heated clouds inside the remnant. The models depends on the traditional
$E$ (the explosion energy), $\rho$ (the preshock ISM density), $t$
(the age of the remnant) plus two parameters describing the evaporating
clouds, namely $C$ (the ratio of mass in clouds to mass in the intercloud
medium) and $\tau$ (the ratio of the cloud evaporation time scale to
the SNR age). With an appropriate choice of normalized quantities
for the radial profiles (for what concern our study, we will consider
the density normalized to the post-shock density, $\rho/\rho_s$,
the temperature normalized to the shell temperature $T/T_s$, and the
normalized emission measure $EM(\pi R_s^2)/EM(tot)$ see their Figure 4),
they show that the dependence on $E$, $\rho_0$ and $t$ can be masked
out in the presentation of the results.  To ease the comparison with
observations, the authors discuss the asymptotic behavior of their
model when $C$ and $\tau \rightarrow \infty$, in which case the model
only depend on the ratio $C/\tau$. In Fig. \ref{cfrwl91}, we show the
normalized temperature, density and emission measure for the models
with $C/\tau=0$, 1, 2, 3 and 4. To further ease the comparison with
the data, we decided to normalize the temperature and density to their
respective central values, instead of their shell values, like in the
work of \citet{wl91}. In fact, in our observations (and in general in
all the MMSNR observations) the low X-ray surface brightness at the shell
prevents an accurate spectral analysis, so it is particularly difficult
to estimate the shell values. On the other hand, the central regions
are by definition the brightest, and we may get more accurate results.
One more point of concern in the comparison between models and our
two targets is the value of the shell radius, used to normalize the
abscissa of the profiles and the emission measure. \srca\  and \srcb\
have far from circular symmetric morphologies and offset X-ray peaks,
so it is particular difficult to estimate a single shell radius. For
\srca, we have used a radius of $10^\prime$ ($\sim 4.3$ pc at 1.5 kpc
distance), because it is the average distance of the radio shell from
the X-ray peak in the East, North and West of the remnant. For \srcb,
we have used $\sim 6.5^\prime$ ($\sim 13$ pc at 4.5 kpc distance),
which is the distance to the east radio wing.

{    In Fig. \ref{cfrwl91}, we also plot the derived values for the low
and high temperature component of \srca\  and for the single thermal component
of \srcb.}
The
profiles of \srcb\  seems to be steeper then the profiles of \srca. In
general, a reasonably good agreement can be found between the observed
normalized emission measure and the model with $C/\tau=2-3$ for both
objects, even if the uncertainties about the adopted shell radius
may affect the actual result. However, it is important to stress that the
density and temperature profiles are not entirely consistent with the
emission measure profiles. In case of \srca, the density profile would
be more in agreement with $C/\tau\ge 4$, while the temperature profile
with $C/\tau < 1$.  {    It is worth noting that the inconsistency
is present in both the thermal components of \srca, and that it also
holds for the density of \srcb\, which would indicate $C/\tau > 4$.}
Conversely, we could use the density profile to
find a good fit to the model by varying the shell radius within a $\pm
50\%$ range, but the emission measure and temperature profile would be
still inconsistent. For instance, we have verified that the emission
measure would be much more centrally peaked than observed.  Finally,
we note that the central peak in both the observed density profiles
seems to exclude evaporating cloud models with low values of $\tau$
(cfr. Fig. 1 in \citealt{wl91}).

In summary, the comparison with the evaporating cloud model points toward
high values of $\tau$, but cannot be considered very informative on the
actual value of $C$ and $\tau$, and indicates that the observed density
profiles are usually steeper than what predicted by the model which best
fit the emission measure profiles, a situation which cannot be explained
in terms of an ejecta component in addition to the evaporating cloud
component. In the latter case, in fact, we would expect the density
profile be shallower than the model, because of the additional density
provided by the ejecta.  We therefore conclude that the X-ray
morphology of these remnants is not compatible with the evaporating
cloud model.  The temperature profiles, though not entirely consistent
with the model, seems to indicate in any case very limited variations,
and therefore points toward efficient thermal conduction inside both
the remnants.  The failure of the cloud evaporation model and the high
efficiency of the thermal conduction suggest an explanation in terms of
the entropy-mixed thermally conductive model suggested by \citet{skp04}
to explain the MMSNR W44, possibly amended to take into account the
peculiar environment in which these remnants are expanding.  A detailed
numerical model of the metal rich MMSNRs taking into account the mixing between
ejecta and shocked ISM material, the thermal conduction and a realistic
model for the environment, is required to verify this scenario.

The comparison with the radiative model of \citet{csm99} can be done
more straightforwardly following \citet{ls06} assuming that the actual
shock radius is the radius at which the remnant enters the radiative
stage. This yields a lower limit of the ambient and central densities,
following the relations of \citet{csm99}. Using realistic radius values,
as the one we have used in the comparison with the cloudy ISM model,
we get ambient density lower limits of the order of 10 cm$^{-3}$. While
it is true that in both the remnants there is substantial evidence of
expansion in dense environments, it is also true that it is very unlikely
that the expansion occurred {\em always} in a medium so dense. For \srca,
\citet{tbr06} argued that the remnant evolved inside wind blown bubble,
an hypothesis originally suggested by \citet{bs86}, while for \srcb,
\citet{gae98} argued the remnant has expanded in a low density hot tunnel before
to encounter recently a high density region. The hot tunnel putative
location is right where the X-ray emission peak is located. In both
cases, therefore, it is very unlikely that a radiative model which
assumes a long evolution inside a very dense medium is applicable,
as it is whenever there is evidence of a massive progenitor star,
which usually have strong pre-supernova winds. \citet{css08} suggest
that the X-ray emission of Kes 27 MMSNR may be explained in term of
a shock reflected by a cavity wall, a mechanism which may be at work
also in \srca\  and \srcb, given the evidences on their environments.

\section{Summary and conclusion}

{    We report on XMM-Newton X-ray observation of the two supernova remnant
\srca\  and \srcb, with particular emphasis on the nature of
their centrally peaked thermal X-ray emission above 1 keV. We confirm that
the X-ray morphology of these remnants entitle them to be included in
the class of mixed morphology SNRs. Contrary to what was previously known,
we find that the chemical abundances of the bright central peak of \srca\
are above solar, promoting this object in the recently pointed out
subclass of metal rich MMSNRs. For \srcb, there is no evidence of enhanced
metal abundances, except maybe for sulfur in the outer regions, but the
limited quality of the data prevented us to perform spatially resolved
spectroscopy of the central peak. 

We derived profiles of metallicity,
temperature, density and emission measure, and we compared them with
prediction of the \citet{wl91} and \citet{csm99} models,
traditionally used to explain the MMSNR morphology in other objects.
We have found that the observed profiles are inconsistent with the models,
and we argue that a more detailed modeling including the mixing of ISM
and ejecta material is required to explain the observations. The entropy
mixing model of \citet{skp04} seems to be the most promising starting
point for a more detailed modeling, though the effects of the complex
environment in which the MMSNRs seems to be located (\citealt{ls06})
and the projection effects as discussed by \citet{pet01} should be taken
into account.  Temperatures, density, and chemical abundances profiles
of MMSNRs are therefore the key observational
constraints to compare with future models. }

We thank D. Leahy for providing us with the 1420 MHz total intensity map
of IC443 in electronic format. We also thank O. Petruk and S. Orlando for
useful comments on this work. This work makes use of results produced by
the PI2S2 Project managed by the Consorzio COMETA, a project co-funded
by the Italian Ministry of University and Research (MIUR) within the
Piano Operativo Nazionale ``Ricerca Scientifica, Sviluppo Tecnologico,
Alta Formazione'' (PON 2000-2006). More information is available at
http://www.consorzio-cometa.it.

\bibliographystyle{aa}
\bibliography{references}

\end{document}